\documentclass[a4paper]{jpconf}

\usepackage{graphicx}
\begin{document}

%
%
\def\d0{D\O}
\def\D0{D\O}
\def\w{$W$}
\def\W{$W$}
\def\z{$Z$}
\def\Z{$Z$}
\def\wg{$W\gamma$}
\def\ww{$WW$}
\def\wpwm{$W^+W^-$}
\def\wz{$WZ$}
\def\wwg{$WW\gamma$}
\def\wwv{$WWV$}
\def\wwz{$WWZ$}
\def\zg{$Z\gamma$}
\def\zzg{$ZZ\gamma$}
\def\zgg{$Z\gamma\gamma$}
\def\pt{$p_T$}
\def\et{$E_T$}
\newcommand{\Eslash}{\mbox{$E \kern-0.6em\slash$}}
\newcommand{\etmis}{\mbox{$\Eslash_T$}}
\def\pdf    {parton distribution function }
\def\pdfs   {parton distribution functions }
\def\ifmath#1{\relax\ifmmode #1\else $#1$}%
\def\TeV{\ifmmode {\mathrm{ Te\kern -0.1em V}}\else
                   \textrm{Te\kern -0.1em V}\fi}%
\def\GeV{\ifmmode {\mathrm{ Ge\kern -0.1em V}}\else
                   \textrm{Ge\kern -0.1em V}\fi}%
\def\MeV{\ifmmode {\mathrm{ Me\kern -0.1em V}}\else
                   \textrm{Me\kern -0.1em V}\fi}%
\def\GeVcc{\ifmmode {\mathrm{ \GeV/c^2}}\else
                   \textrm{Ge\kern -0.1em V/c$^2$}\fi}%
\def\MeVcc{\ifmmode {\mathrm{ \MeV/c^2}}\else
                   \textrm{Me\kern -0.1em V/c$^2$}\fi}%
\def\pbar               {\mbox{$\overline{p}$}}
\def\pbarp              {\mbox{$\overline{p}p$}}
\def\ppbar              {\mbox{$p\overline{p}$}}
\def\ttbar              {\mbox{$t\overline{t}$}}
\newcommand{\Afb}       {A_{\mathrm{FB}}}
\newcommand{\AFB}       {A_{\mathrm{FB}}}
\newcommand{\ee}        {\mbox{$e^+e^-$}}
\newcommand{\qq}        {\mbox{$q\overline q$}}
\newcommand{\uu}        {\mbox{$u\overline u$}}
\newcommand{\dd}        {\mbox{$d\overline d$}}
\newcommand{\bb}        {\mbox{$b\overline b$}}
\newcommand{\SM}        {\mbox{SM}}
\def\eg{{\it e.g.}}
\def\ie{{\it i.e.}}
\def\etal{{\it et al.}}
\def\isajet{{\sc isajet}}
\def\geant{{\sc geant}}
\def\pythia{{\sc pythia}}
\def\vecbos{{\sc vecbos}}
\def\herwig{{\sc herwig}}
\newcommand{\tab}{\hspace*{0.5in}}

\def\grad{\nabla}
\def\gradvec{\mbox{\boldmath $\nabla$}}
\def\Aslash{\mbox{${\hbox{$A$\kern-0.55em\hbox{/}}}$}}
\def\pslash{\mbox{${\hbox{$p$\kern-0.45em\hbox{/}}}$}}


%
%
\def\andit{{\it\&}}
\def\half{\frac{1}{2}}
\def\thalf{\tfrac{1}{2}} 
\def\third{\frac{1}{3}}
\def\quarter{\frac{1}{4}}
\def\rii{\sqrt{2}}
\def\to{\rightarrow}
\def\S{\mathhexbox279}
\def\gesim{\,{\raise-3pt\hbox{$\sim$}}\!\!\!\!\!{\raise2pt\hbox{$>$}}\,}
\def\lesim{\,{\raise-3pt\hbox{$\sim$}}\!\!\!\!\!{\raise2pt\hbox{$<$}}\,}
\def\boldoverdot{\,{\raise6pt\hbox{\bf.}\!\!\!\!\>}}
\def\re{{\bf Re}}
\def\im{{\bf Im}}
\def\ie{{i.e.}}
\def\cf{{\it cf.}\ }
\def\ibid{{\it ibid.}\ }
\def\etal{{\it et. al.}}
\def\acal{{\cal A}}
\def\bcal{{\cal B}}
\def\ccal{{\cal C}}
\def\dcal{{\cal D}}
\def\ecal{{\cal E}}
\def\fcal{{\cal F}}
\def\gcal{{\cal G}}
\def\hcal{{\cal H}}
\def\ical{{\cal I}}
\def\jcal{{\cal J}}
\def\kcal{{\cal K}}
\def\lcal{{\cal L}}
\def\mcal{{\cal M}}
\def\ncal{{\cal N}}
\def\ocal{{\cal O}}
\def\pcal{{\cal P}}
\def\qcal{{\cal Q}}
\def\rcal{{\cal R}}
\def\scal{{\cal S}}
\def\tcal{{\cal T}}
\def\ucal{{\cal U}}
\def\vcal{{\cal V}}
\def\wcal{{\cal W}}
\def\xcal{{\cal X}}
\def\ycal{{\cal Y}}
\def\zcal{{\cal Z}}
\def\iBB{ \hbox{{\mysmallii I}}\!\hbox{{\mysmallii I}} }
\def\bBB{ \hbox{{\mysmallii I}}\!\hbox{{\mysmallii B}} }
\def\pBB{ \hbox{{\mysmallii I}}\!\hbox{{\mysmallii P}} }
\def\rBB{ \hbox{{\mysmallii I}}\!\hbox{{\mysmallii R}} }
\def\alpbf{{\pmb{$\alpha$}}}
\def\betbf{{\pmb{$\beta$}}}
\def\gambf{{\pmb{$\gamma$}}}
\def\delbf{{\pmb{$\delta$}}}
\def\epsbf{{\pmb{$\epsilon$}}}
\def\zetbf{{\pmb{$\zeta$}}}
\def\etabf{{\pmb{$\eta$}}}
\def\thebf{{\pmb{$\theta$}}}
\def\varthebf{{\pmb{$\vartheta$}}}
\def\iotbf{{\pmb{$\iota$}}}
\def\kapbf{{\pmb{$\kappa$}}}
\def\lambf{{\pmb{$\lambda$}}}
\def\mubf{{\pmb{$\mu$}}}
\def\nubf{{\pmb{$\nu$}}}
\def\xibf{{\pmb{$\xi$}}}
\def\pibf{{\pmb{$\pi$}}}
\def\varpibf{{\pmb{$\varpi$}}}
\def\rhobf{{\pmb{$\rho$}}}
\def\sigbf{{\pmb{$\sigma$}}}
\def\taubf{{\pmb{$\tau$}}}
\def\upsbf{{\pmb{$\upsilon$}}}
\def\phibf{{\pmb{$\phi$}}}
\def\varphibf{{\pmb{$\varphi$}}}
\def\chibf{{\pmb{$\chi$}}}
\def\psibf{{\pmb{$\psi$}}}
\def\omebf{{\pmb{$\omega$}}}
\def\Gambf{{\pmb{$\Gamma$}}}
\def\Delbf{{\pmb{$\Delta$}}}
\def\Thebf{{\pmb{$\Theta$}}}
\def\Lambf{{\pmb{$\Lambda$}}}
\def\Xibf{{\pmb{$\Xi$}}}
\def\Pibf{{\pmb{$\Pi$}}}
\def\Sigbf{{\pmb{$\sigma$}}}
\def\Upsbf{{\pmb{$\Upsilon$}}}
\def\Phibf{{\pmb{$\Phi$}}}
\def\Psibf{{\pmb{$\Psi$}}}
\def\Omebf{{\pmb{$\Omega$}}}
\def\ssb{spontaneous symmetry breaking}
\def\vev{vacuum expectation value}
\def\irrep{irreducible representation}
\def\lhs{left hand side\ }
\def\rhs{right hand side\ }
\def\Ssb{Spontaneous symmetry breaking\ }
\def\Vev{Vacuum expectation value}
\def\Irrep{Irreducible representation}
\def\Lhs{Left hand side\ }
\def\Rhs{Right hand side\ }
\def\tr{ \hbox{tr}}
\def\det{\hbox{det}}
\def\Tr{ \hbox{Tr}}
\def\Det{\hbox{Det}}
\def\diag{\hbox{\diag}}
\def\sm{Standard Model}
\def\ev{\hbox{eV}}
\def\kev{\hbox{keV}}
\def\mev{\hbox{MeV}}
\def\gev{\hbox{GeV}}
\def\tev{\hbox{TeV}}
\def\milm{\hbox{mm}}
\def\cm{\hbox{cm}}
\def\m{\hbox{m}}
\def\km{\hbox{km}}
\def\gr{\hbox{gr}}
\def\kg{\hbox{kg}}
%
%
\def\noteeye{ {{$\quad(\!(\subset\!\!\!\!\bullet\!\!\!\!\supset)\!)\quad$}}}
\def\note#1{{\bf \noteeye\nobreak #1 \noteeye } }
\def\doubleundertext#1{
{\undertext{\vphantom{y}#1}}\par\nobreak\vskip-\the\baselineskip\vskip4pt%
\undertext{\hbox to 2in{}}}
\def\inbox#1{\vbox{\hrule\hbox{\vrule\kern5pt
     \vbox{\kern5pt#1\kern5pt}\kern5pt\vrule}\hrule}}
\def\sqr#1#2{{\vcenter{\hrule height.#2pt
      \hbox{\vrule width.#2pt height#1pt \kern#1pt
         \vrule width.#2pt}
      \hrule height.#2pt}}}
\def\today{\ifcase\month\or
  January\or February\or March\or April\or May\or June\or
  July\or August\or September\or October\or November\or December\fi
  \space\number\day, \number\year}
\def\pmb#1{\setbox0=\hbox{#1}%
  \kern-.025em\copy0\kern-\wd0
  \kern.05em\copy0\kern-\wd0
  \kern-.025em\raise.0433em\box0 }
\def\up#1{^{\left( #1 \right) }}
\def\lowti#1{_{{\rm #1 }}}
\def\inv#1{{\frac{1}{#1}}}
\def\deriva#1#2#3{\left({\frac{\partial #1}{\partial #2}}\right)_{#3}}
\def\su#1{{SU(#1)}}
\def\ui{U(1)}
\def\antes{}
\def\despues{.}
\def\dss{ {}^2 }
%
\def\sumprime_#1{\setbox0=\hbox{$\scriptstyle{#1}$}
  \setbox2=\hbox{$\displaystyle{\sum}$}
  \setbox4=\hbox{${}'\mathsurround=0pt$}
  \dimen0=.5\wd0 \advance\dimen0 by-.5\wd2
  \ifdim\dimen0>0pt
  \ifdim\dimen0>\wd4 \kern\wd4 \else\kern\dimen0\fi\fi
\mathop{{\sum}'}_{\kern-\wd4 #1}}
\def\dodraft#1{
\typeout{}
\typeout{**************|||||||||******************}
\typeout{}
\typeout{Paper: #1 (DRAFT)}
\typeout{}
\typeout{**************|||||||||******************}
\typeout{}
\typeout{}
}
%
%
\font\sanser=cmssq8
\font\sanseru=cmssq8 scaled\magstep1 
\font\sanserd=cmssq8 scaled\magstep2 
\font\sanseri=cmssq8 scaled\magstep1
\font\sanserii=cmssq8 scaled\magstep2
\font\sanseriii=cmssq8 scaled\magstep3
\font\sanseriv=cmssq8 scaled\magstep4
\font\sanserv=cmssq8 scaled\magstep5
\font\tyt=cmtt10
\font\tyti=cmtt10 scaled\magstep1
\font\tytii=cmtt10 scaled\magstep2
\font\tytiii=cmtt10 scaled\magstep3
\font\tytiv=cmtt10 scaled\magstep4
\font\tytv=cmtt10 scaled\magstep5
\font\slanti=cmsl10 scaled\magstep1
\font\slantii=cmsl10 scaled\magstep2
\font\slantiii=cmsl10 scaled\magstep3
\font\slantiv=cmsl10 scaled\magstep4
\font\slantv=cmsl10 scaled\magstep5
\font\bigboldi=cmbx10 scaled\magstep1
\font\bigboldii=cmbx10 scaled\magstep2
\font\bigboldiii=cmbx10 scaled\magstep3
\font\bigboldiv=cmbx10 scaled\magstep4
\font\bigboldv=cmbx10 scaled\magstep5
\font\mysmall=cmr8
\font\mysmalli=cmr8 scaled\magstep1
\font\mysmallii=cmr8 scaled\magstep2
\font\mysmalliii=cmr8 scaled\magstep3
\font\mysmalliv=cmr8 scaled\magstep4
\font\mysmallv=cmr8 scaled\magstep5
\font\ital=cmti10
\font\itali=cmti10 scaled\magstep1
\font\italii=cmti10 scaled\magstep2
\font\italiii=cmti10 scaled\magstep3
\font\italiv=cmti10 scaled\magstep4
\font\italv=cmti10 scaled\magstep5
\font\smallit=cmmi7
\font\smalliti=cmmi7 scaled\magstep1
\font\smallitii=cmmi7 scaled\magstep2
\font\rmi=cmr10 scaled\magstep1
\font\rmii=cmr10 scaled\magstep2
\font\rmiii=cmr10 scaled\magstep3
\font\rmiv=cmr10 scaled\magstep4
\font\rmv=cmr10 scaled\magstep5
\font\eightrm=cmr8

\title{B-Physics at the Tevatron}

\author{John Ellison\\ (for the CDF and \d0\ Collaborations)}

\address{Department of Physics and Astronomy, University of California, Riverside CA 92521 USA}

\ead{john.ellison@ucr.edu}

\begin{abstract}
Recent $B$-Physics results from the CDF and \d0\ experiments at the Tevatron are described, with emphasis on rare decays and searches for CP violation.
\end{abstract}

\section{Introduction}
We report on recent $B$-Physics results from the Tevatron. The topics covered include measurement of the polarization amplitudes in $B_s^0 \to \phi \phi$, the search for rare flavor-changing neutral-current decays, CP violation in $B_s^0 \to J/\psi \phi$ and semileptonic $B_s^0$ decays, and a new measurement of the like-sign asymmetry in dimuon events. 

The data used for the results described here are based on approximately $2.9 - 6.1$~fb$^{-1}$ of data collected by the CDF and \d0\ experiments at the Fermilab Tevatron, a proton-antiproton collider operating at a center-of-mass energy of 1.96~TeV. The \d0\ and CDF detectors are described in detail in \cite{d0det} and \cite{cdfdet} respectively.

\section{Polarization Amplitudes in $B_s^0 \to \phi \phi$}

The decay $B_s^0 \to \phi \phi$ is an example of a pseudoscalar meson decaying to two vector mesons and is described by three polarization amplitudes: $A_0, A_\parallel$, and $A_\perp$ corresponding to the three polarization configurations of the final state decay products. Due to the $V-A$ nature of the weak interaction and helicity conservation in QCD,  the naive expectation is that the longitudinal polarization amplitude dominates in  $B$ decays to two light vector mesons:  $| A_0 |^2 \gg | A_\parallel |^2 \approx | A_\perp |^2$. This has been confirmed in tree-level $b \to u$ decays and there is evidence for this in $d \to d$ penguin decays. However, in the decay $B \to \phi K^*$, a $b \to s$ decay, it has recently been found that $| A_0 |^2 \approx | A_\parallel |^2 \approx | A_\perp |^2$. Since the $B_s^0 \to \phi \phi$ decay also proceeds through a penguin decay, it can be used to check this result. Also, with sufficient statistics, this decay channel will be useful for a comparison of CP violation in  $B_s^0 \to \phi \phi$ and $B_s^0 \to J/\psi \phi$
decays. 

CDF have studied the $B_s^0 \to \phi \phi$ decay, where both $\phi$ mesons decay via $\phi \to K^+ K^-$, using 300 events reconstructed in 2.9~fb$^{-1}$ of data~\cite{cdf_Bsphiphi}. The polarization amplitudes are measured using a fit to the mass and three decay angles of the $B_s^0$ decay products. The fit results are:
\begin{eqnarray*}
| A_0 |^2  &=& 0.348 \pm 0.041 {\rm (stat)} \pm 0.021 {\rm (syst)}  \\
| A_\parallel |^2 &=& 0.287 \pm 0.043 {\rm (stat)} \pm 0.011 {\rm (syst)}  \\
| A_\perp |^2 &=& 0.365 \pm 0.044 {\rm (stat)} \pm 0.027 {\rm (syst)} \\
\cos{\delta_\parallel} &=&  -0.91 \thinspace ^{+0.15}_{-0.13} {\rm (stat)}
\pm 0.09 {\rm (syst)}
\end{eqnarray*}
 
\noindent
where $\cos{\delta_\parallel}$ is the relative strong phase. These results show that $| A_0 |^2 \approx | A_\parallel |^2 \approx | A_\perp |^2$  and therefore the naive prediction is disfavored. The results are consistent within the uncertainties with the theoretical expectations of QCD factorization calculations (see \cite{Beneke} for example), while they do not agree with simple perturbative QCD \cite{Ali}.

\section{Rare Decays: $B_{s,d} \to \mu^+ \mu^-$}

The decays $B_{s,d} \to \mu^+ \mu^-$ are flavor changing neutral current processes that proceed via \mbox{GIM-suppressed} penguin diagrams, and consequently they have small branching fractions in the SM \cite{Buras}:
\begin{eqnarray*}
B(B_s^0 \to \mu^+ \mu^-) &=& (3.6 \pm 0.3) \times 10^{-9} \\
B(B^0 \to \mu^+ \mu^-) &=& (1.1 \pm 0.1) \times 10^{-10}
\end{eqnarray*}

\noindent
The SM signal is currently beyond the sensitivity of CDF and \d0\ at the Tevatron, but new physics can significantly enhance the branching fractions. Examples are the MSSM ($B \sim \tan^6{\beta}$), SUSY R-parity violating models, and GUTs. Therefore, the observation of these decays with the current data would necessarily indicate new physics.

\d0\ have reported a new result in the search for $B_s^0 \to \mu^+ \mu^-$ based on 6.1~fb$^{-1}$ of data~\cite{d0_Bsmumu}. Events are selected using a Bayesian neural network based on five variables:  
the $B_s^0$ transverse momentum and pointing angle; the vertex $\chi^2$ and transverse decay length significance; and the smallest impact parameter significance of the two muons. A control sample of $B^+ \to J/\psi [\to \mu^+ \mu^-] K^+$ is used for normalization. Figure~\ref{fig:Bsmumu} 
shows the $\mu \mu$ invariant mass distribution for events in the signal region of the neural network discriminant $\beta$. 
\begin{figure}[htp]
	\begin{center}
	\includegraphics[width=0.6\textwidth]{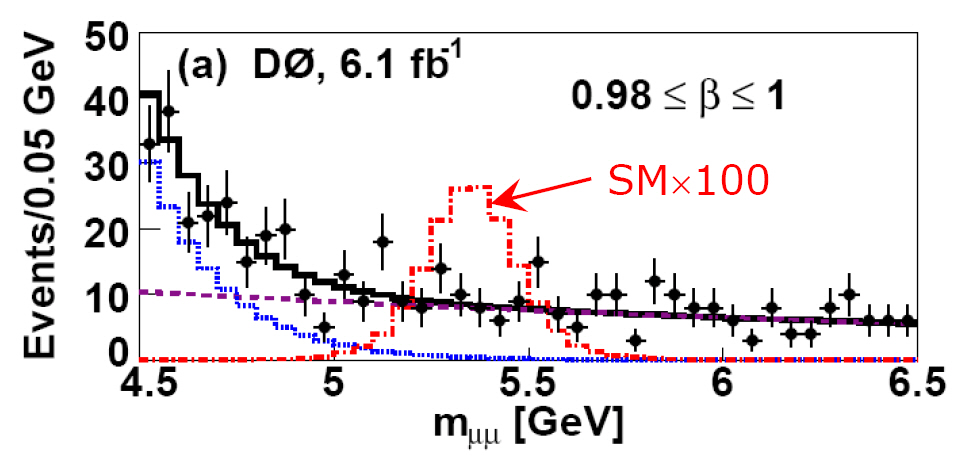}
	\caption{\label{fig:Bsmumu}
	The $\mu \mu$ invariant mass distribution for events in the signal region of the neural 
	network discriminant $\beta$, showing the data (dots with uncertainties),
  expected background distribution (solid line), and the SM
  signal distribution multiplied by a factor of 100. The dimuon background contributions from 
  double semileptonic decays $B(D) \to \mu^+ \nu X$ (dashed line) and sequential semileptonic decays $B \to \mu^+ \nu \bar D, \bar D \to \mu^- \bar \nu X$ (dotted line) are also
  shown.}
	\end{center}
\end{figure}

No evidence for a signal is observed and a limit is set on the branching fraction at the 95\% C.L.:
\begin{eqnarray*}
B(B_s^0 \to \mu^+ \mu^-) < 5.1 \times 10^{-8}
\end{eqnarray*}

\noindent
with an expected limit of $3.8 \times 10^{-8}$.

CDF have searched for $B_s^0 \to \mu^+ \mu^-$ and $B^0 \to \mu^+ \mu^-$ in a data set of 3.7~fb$^{-1}$ and set the following 95\% C.L. limits~\cite{cdf_Bsmumu}:
\begin{eqnarray*}
B(B_s^0 \to \mu^+ \mu^-) < 4.3 \times 10^{-8} \\
B(B^0 \to \mu^+ \mu^-) < 7.6 \times 10^{-9} 
\end{eqnarray*}
\noindent
with expected limits of $3.3 \times 10^{-8}$ and $9.1 \times 10^{-9}$, respectively.

\section{CP Violation in $B_s^0 \to J/\psi \phi$}

The $B_s^0 \to J/\psi \; \phi$ decay involves a final state that is a mixture of CP-even and CP-odd states. CP violation can occur via the interference between direct and mixed decays and is described by the phase $\beta_s$. CDF have updated their analysis of this mode using a 5.2~fb$^{-1}$ data 
set~\cite{cdf_betas}. In order to separate the CP-even and CP-odd states, CDF use a maximum likelihood fit to the mass, lifetime, and time-dependent angular distributions of the $B_s^0 \to J/\psi (\to \mu^+ \mu^-) \; \phi (K^+ K^-)$ decay. The fit yields the CP-violating phase $\beta_s$ and the width difference $\Delta \Gamma_s \equiv \Gamma_L - \Gamma_H$, where $\Gamma_L$ and $\Gamma_H$ are the widths of the light and heavy eigenstates. 

The $B_s^0$ flavor at production is determined using a combined opposite-side plus same­side tagging algorithm. The initial state flavor tag improves the sensitivity to the CP-violating phase and removes a sign ambiguity on $\beta_s$ for a given $\Delta \Gamma_s$. In the fit, $\Delta M_s$ is constrained to its measured value. The number of signal events reconstructed is 6500.
The average lifetime and width difference resulting form the fit are:
\begin{eqnarray*}
   \bar \tau_s &=& 1.53 \pm 0.025 {\rm (stat)} \pm 0.012 {\rm (syst)}~\rm{ps} \\ 
   \Delta \Gamma_s  &=& 0.075 \pm 0.035 {\rm (stat)} 
      \pm 0.01{\rm (syst)}~\rm{ps}^{-1}
\end{eqnarray*}

\noindent
The likelihood is coverage-corrected to account for non-Gaussian tails and systematic uncertainties. Confidence level contours in the $\phi_s - \Delta \Gamma_s$ plane are shown in Fig.~\ref{fig:dg_betas}(a). The likelihood as a function of $\beta_s$ is shown in Fig.~\ref{fig:dg_betas}(b). The resulting 95\% C.L. limits on the CP-violating phase $\beta_s$ are:
\begin{eqnarray*}
  \beta_{s} = [-\pi/2, -1.44] ~ \cup ~ [-0.13, 0.68] ~ \cup ~ [0.89, \pi/2] ~~~~ 95\% ~ \mathrm{C.L.}
\end{eqnarray*}
\begin{figure}[htp]
\begin{center}
\begin{tabular}{c c}
	\includegraphics[width=0.45\textwidth]{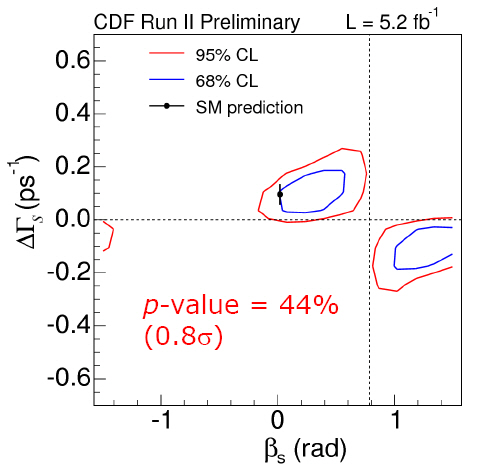}
	\includegraphics[width=0.45\textwidth]{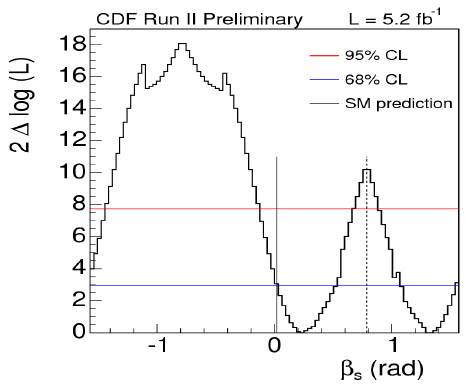}
\end{tabular}
\end{center}
	\caption{\label{fig:dg_betas}
	(a) 65\% and 95\% confidence level regions in the $\beta_s - \Delta \Gamma_s$ plane for the CDF 
	$B_s^0 \to J/\psi \phi$ analysis based on 5.2~fb$^{-1}$ of data. (b) Coverage-corrected 
	profile likelihood as a function of $\beta_s$.}
\end{figure}

\noindent
The $p$-value for the fit is 44\%, corresponding to a deviation from the standard model prediction of 0.8~$\sigma$.

\section{Search for CP Violation in Semileptonic $B_s^0$ Decays}

In this section we report a new search for CP violation in the decay 
$B_s^0  \to D_s^-  \mu^+ \nu X$ by measurement of the flavor-specific asymmetry using a time-dependent analysis with flavor tagging.  Two final state samples were reconstructed:
$D_s^- \to \phi \pi^- , \;\; \phi \to K^+ K^-$ and
$D_s^- \to K^{*0} K^- , \;\; K^{*0} \to K^+ \pi^-$.

The flavor-specific asymmetry is defined as
\begin{eqnarray*}
	a_{fs}^s = \frac{\Gamma \left( \bar B_s^0(t) \to f \right) - \Gamma \left( B_s^0(t) \to \bar f \right)}{\Gamma \left( \bar B_s^0(t) \to f \right) + \Gamma \left( B_s^0(t) \to \bar f \right)} 
\end{eqnarray*}

\noindent
The SM prediction is very small \cite{Lenz}, $a_{fs}^s = (-2.06 \pm 0.57) \times 10^{-5}$, and any deviation from this prediction would be evidence for CP-violating new physics.

The analysis uses a data set of integrated luminosity 5~fb$^{-1}$~\cite{d0_semilep}. The technique used is
similar to that used in the \d0\ $B_s^0$ oscillation analysis. 
A fit to the invariant $KK\pi$ mass of the selected data is shown in Fig.~\ref{fig:m_KKpi}.
\begin{figure}[htp]
	\begin{center}
	\includegraphics[width=0.95\textwidth]{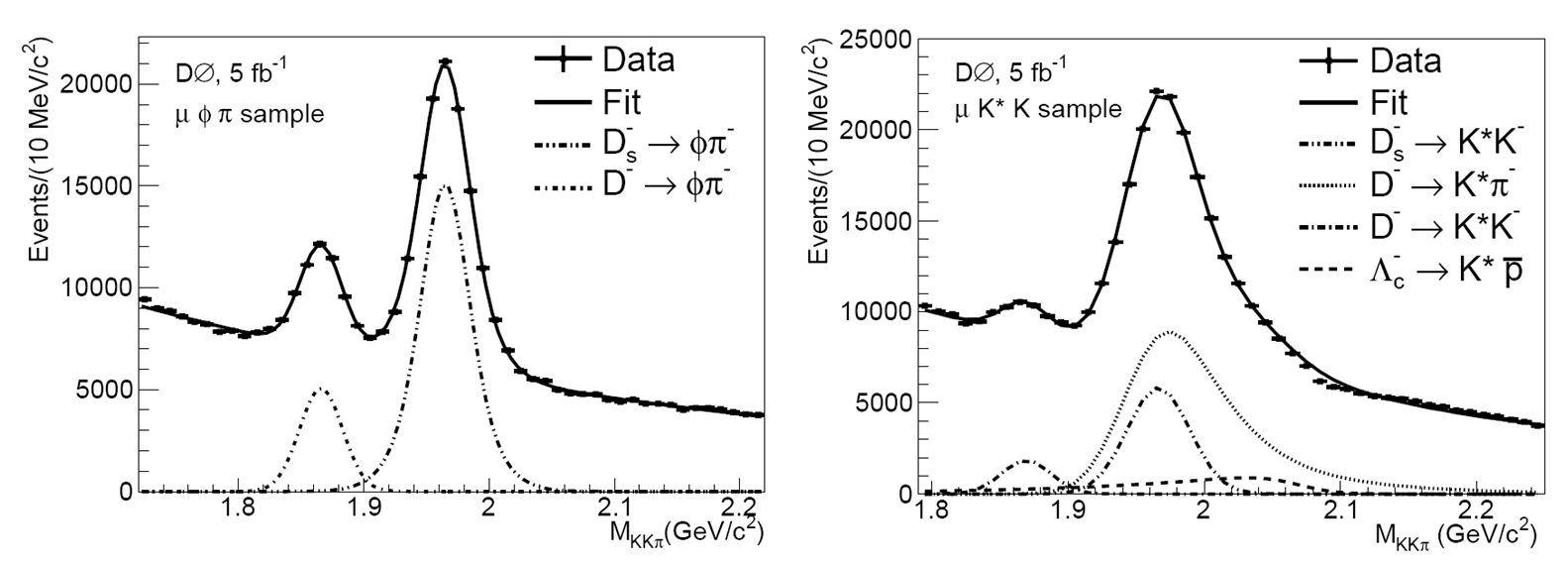}
	\caption{\label{fig:m_KKpi}
	The $KK \pi$ invariant mass distributions in the $\mu \phi \pi$ and $\mu K^{*0} K$ samples
	together with the fit results.}
	\end{center}
\end{figure}

An unbinned likelihood fit is used to extract the asymmetry. The systematic uncertainties are mainly due to uncertainties in the $c \bar c$ contribution, uncertainties in the efficiency vs visible proper decay length, and uncertainties in the $B_s^0 \to D_s^{(*)} \mu \nu$ branching fractions. Accounting for these yields the final result for the flavor-specific asymmetry: 
\begin{eqnarray*}
	a_{fs}^s  =  - 0.0017 \pm 0.0091 \rm{(stat)} \thinspace ^{+0.0014}_{-0.0015} \rm{(syst)}
\end{eqnarray*}

This result is consistent with the SM prediction and is the most precise measurement to date.

\section{Anomalous Like-Sign Dimuon Asymmetry}

The final result reported here is the new \d0\ measurement of the like-sign dimuon 
asymmetry~\cite{d0_dimuon}. The dimuon charge asymmetry of semileptonic $B$ decays is defined by
\begin{eqnarray*}
	A_{sl}^b  \equiv  \frac{N_b^{++} - N_b^{--}}{N_b^{++} + N_b^{--}}
\end{eqnarray*}

\noindent
and has contributions from $B_d$ and $B_s^0$ decays:
\begin{eqnarray*}
  A_{sl}^b  =  (0.506 \pm 0.043) a_{sl}^d +  (0.494 \pm 0.043) a_{sl}^s
\end{eqnarray*}
\noindent where
\begin{eqnarray*}
	a_{sl}^q  =  \frac{\Gamma(\bar B^0_q \to \mu^+ X) - \Gamma(B^0_q \to \mu^- X)}{\Gamma(\bar B^0_q \to \mu^+ X) + \Gamma(B^0_q \to \mu^- X)}, ~~~~~~~~~ q = d,s \\	
\end{eqnarray*}

\noindent
The SM prediction for the semileptonic charge asymmetry is very small \cite{Lenz},
$A_{sl}^b = (-2.3  \thinspace ^{+0.5}_{-0.6}) \times 10^{-4}$, and new physics could change its value if new CP-violating phases are introduced.

The ``raw'' asymmetries measured by \d0\ are the like-sign dimuon asymmetry
\begin{eqnarray*}
	A  \equiv  \frac{N(\mu^+ \mu^+) - N(\mu^- \mu^-)}{N(\mu^+ \mu^+) + N(\mu^- \mu^-)}
	   = (0.564 \pm 0.053)\%
\end{eqnarray*}
\noindent and the inclusive muon charge asymmetry 
\begin{eqnarray*}
	a  \equiv  \frac{N(\mu^+) - N(\mu^-)}{N(\mu^+) + N(\mu^-)}
     = (0.955 \pm 0.003)\%
\end{eqnarray*}

\noindent
These measurements are obtained from 3.7 million events in the like-sign dimuon sample and 1.5 billion events in the inclusive muon sample, respectively. The relation of $A_{sl}^b$ to these raw asymmetries is as follows:
\begin{eqnarray*}
	A  &=&  K A_{sl}^b + A_{\rm{bkg}} \\
	a  &=&  k A_{sl}^b + a_{\rm{bkg}}
\end{eqnarray*}

After accounting for non-symmetric background processes from "prompt/physics" sources ($A_{\rm{bkg}}$ and $a_{\rm{bkg}}$) and dilution due to other processes that do not contribute to asymmetry ($K$ and $k$), the final result is:
\begin{eqnarray*}
	A_{sl}^b  =  (-0.957 \pm 0.251 {\rm (stat)} \pm 0.146 {\rm (syst))} \%
\end{eqnarray*}

\noindent
which represents a 3.2~$\sigma$ deviation from the theoretical prediction of the SM. A comparison with other measurements is shown in Fig.~\ref{fig:asld_vs_asls}. As can be seen, the result is consistent with other measurements with the uncertainties.
\begin{figure}[htp]
	\begin{center}
	\includegraphics[width=0.6\textwidth]{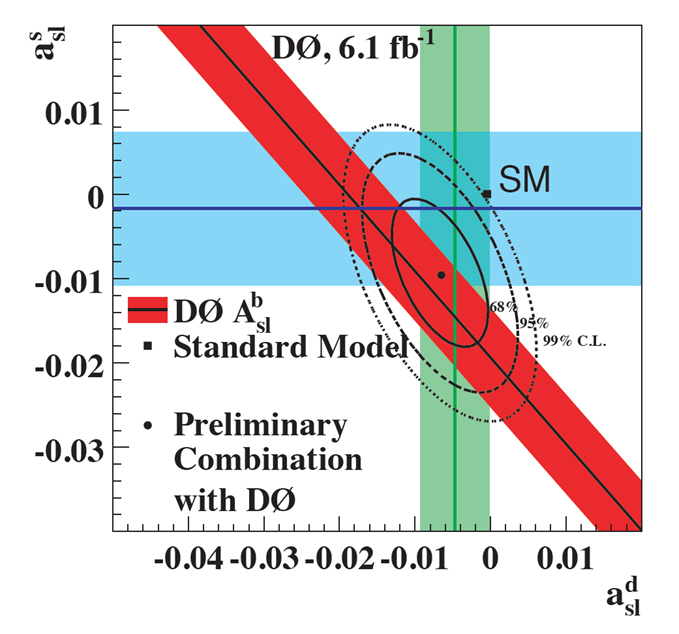}
	\caption{\label{fig:asld_vs_asls}
  $a_{sl}^d$ vs. $a_{sl}^s$ showing the \d0\ measurement of $A_{sl}^b$ (black line with red 
  error  band), the world average measurement of $a_{sl}^d$ (green line with green error band), 
  and the \d0\ direct measurement of $a_{sl}^s$ (blue line with blue error band). Also shown are 
  the 68\%, 95\%, and 99\% C.L. regions for the combination of the three results.}
	\end{center}
\end{figure}

\section{Conclusions}
The high luminosity of the Tevatron coupled with the relatively large $b$ production cross section allows many measurements to be made that provide tests of the SM and allow searches for physics beyond the SM, especially in the area of CP violation and rare decays. The measurements are ongoing and are complementary to results obtained at the B-factories. Future measurements at the Tevatron and at the LHC will further explore the rich field of flavor physics.

\end{document}